\documentclass{article}

\usepackage[pdftex]{graphicx}
\graphicspath{{./graphics/}}
\DeclareGraphicsExtensions{.pdf}

\usepackage{amsmath}
\usepackage{amssymb}
\usepackage{algorithmic}
\usepackage{array}
\usepackage{color}
\usepackage[hidelinks]{hyperref}
\usepackage[tableposition=top]{caption}
\usepackage{subcaption}
\usepackage{cite}
\usepackage[table,xcdraw]{xcolor}
\usepackage{spconf}
\usepackage{multirow}
\usepackage[symbol]{footmisc}

\usepackage{enumitem}
\setlist[itemize]{leftmargin=*}
\setlist[enumerate]{leftmargin=*}

\newlength{\bibitemsep}
\newlength{\bibparskip}
\let\oldthebibliography\thebibliography
\renewcommand\thebibliography[1]{%
  \oldthebibliography{#1}%
  \setlength{\parskip}{\bibitemsep}%
  \setlength{\itemsep}{\bibparskip}%
}

\DeclareMathOperator*{\argmin}{arg\,min}

\begin{document}

\title{\vspace{-4mm}On permutation invariant training for speech source separation}

\name{Xiaoyu~Liu \quad Jordi~Pons}

\address{Dolby Laboratories}

\markboth{Submitted to ICASSP 2021}%
{Shell \lowercase{\textit{et al.}}: Title}%
\maketitle
\ninept

\begin{abstract}

\vspace{0.15mm}

\noindent We study permutation invariant training (PIT), which targets at the permutation ambiguity problem for speaker independent source separation models. We extend two state-of-the-art PIT strategies. First, we look at the two-stage speaker separation and tracking algorithm based on frame level PIT (tPIT) and clustering, which was originally proposed for the STFT domain, and we adapt it to work with waveforms and over a learned latent space. Further, we propose an efficient clustering loss scalable to waveform models. Second, we extend a recently proposed auxiliary speaker-ID loss with a deep feature loss based on ``problem agnostic speech features", to reduce the local permutation errors made by the utterance level PIT (uPIT). Our results show that the proposed extensions help reducing permutation ambiguity. However, we also note that the studied STFT-based models are more effective at reducing permutation errors than waveform-based models, a perspective overlooked in recent studies.  

\end{abstract}
\vspace{-2mm}

\section{Introduction}
\vspace{-1mm}

\label{sec:introduction}

The permutation ambiguity problem occurs when training speaker independent deep learning models in a supervised fashion.
The goal of such systems is to separate $C$~speech sources $x_c(t)$ from their mixture waveform $y(t) = \sum_{c=1}^C x_c(t)$.
Accordingly, $C!$ permutations of speaker outputs $\hat{x}_c(t)$ to ground truth $x_c(t)$ pairs exist for computing the loss.
However, these pairs cannot be arbitrarily assigned due to the speaker independent nature of the task. Utterance level permutation invariant training (uPIT)~\cite{kolbaek2017multitalker} addresses this problem by minimizing the smallest separation loss of all permutations computed over the entire utterance, thus enforcing permutation consistency across frames. However, the assumption that all frames share the same permutation may lead to sub-optimal frame level separation, causing local speaker swaps and leakage~\cite{liu2019divide}. On the other hand, frame level PIT (tPIT)~\cite{yu2017permutation} performs PIT for each frame independently, thus achieves excellent frame level separation quality. However, the permutation frequently changes over frames at inference time. Improvements to PIT roughly fall into two categories: (i)~designing a permutation (or speaker) tracking algorithm for tPIT~\cite{liu2019divide, liu2018casa,zeghidour2020wavesplit}; 
and (ii)~designing better uPIT objectives to further strengthen permutation consistency~\cite{xu2018single,yousefi2019probabilistic,yang2020interrupted,nachmani2020voice}.
Along these two lines, our work takes a close look at tPIT+clustering, a recent idea introduced by Deep CASA~\cite{liu2019divide}, that targets at accurate frame level separation (tPIT) and speaker tracking (clustering) in two stages.
We also explore another promising method based on \mbox{uPIT+speaker-ID} loss~\cite{nachmani2020voice}, that introduces an additional deep feature loss term (speaker-ID) to help uPIT reducing local speaker swaps. In this paper, we extend these two training strategies for Conv-TasNet~\cite{luo2019conv}, a fully convolutional version of TasNet~\cite{luo2018tasnet, luo2019conv, luo2020dual, kadiouglu2020empirical} that models speaker separation in the waveform domain.

In section 2, we extend tPIT+clustering training algorithm by Deep CASA, an spectrogram-based model, to Conv-TasNet, which uses very short waveform frames (such as 2 ms). We find that tPIT based on such short waveform frames can be challenging. Therefore, we propose performing tPIT in a pre-trained latent space---which allows for a more meaningful feature space for tPIT than the short waveform frames. Further, when training the clustering model, Deep CASA employs a memory and computationally expensive pairwise similarity loss that does not scale for waveform inputs. We propose a loss that reduces the complexity from quadratic to linear, making the training of the clustering model feasible for waveform models. 

In section 3, we also extend the uPIT+speaker-ID loss with PASE, a problem agnostic speech encoder~\cite{pascual2019learning, ravanelli2020multi, alvarezproblem}. PASE is pre-trained in a self-supervised fashion with a collection of objectives much broader than speaker-ID, to extract general-purpose speech embeddings from waveforms. 
In addition, we also look at conditioning Conv-TasNet with PASE embeddings in a cascaded system consisting of two steps: (i)~uPIT+PASE speaker separation, and (ii)~conditioning Conv-TasNet with PASE embeddings computed from the speakers separated in step (i).

Conv-TasNet and Deep CASA can both be interpreted as architectures with an encoder/decoder and a separator, where the encoder/decoder in Conv-TasNet is learnable whereas in Deep CASA is the STFT. Previous works have already compared learnable and signal processing-based encoders/decoders~\cite{ditter2020multi, heitkaemper2020demystifying, kavalerov2019universal, pariente2020filterbank, bahmaninezhad2019comprehensive}. However, from the permutation ambiguity perspective, it remains an open question if using waveform- or STFT-based models has any advantages. Section 4 shows that our tPIT+clustering and uPIT+PASE extensions for Conv-TasNet help on reducing permutation errors and on generalization. 
However, Deep CASA~outperforms the Conv-TasNet variants we study, highlighting the advantages of STFT-based models and the remaining challenges for waveform-based models from the permutation ambiguity perspective.

\vspace{-2mm}

\section{\lowercase{t}PIT + clustering for conv-TasNet}
\vspace{-1mm}
\label{sec:the_algorithm}

\noindent In this section, we introduce the tPIT+clustering algorithm in Deep CASA, and explain how we adapt it for Conv-TasNet. Deep CASA~\cite{liu2019divide} is formulated for two speakers ($C$$=$$2$), but can be generalized to more speakers~\cite{liu2018casa}. We also set $C$$=$$2$ for our work.

\vspace{-2mm}
\subsection{tPIT step}
\vspace{-0mm}

We investigate several tPIT variants: (i) \textbf{tPIT-STFT}, the tPIT step used in Deep CASA for spectrogram-based models; (ii) \textbf{tPIT-time}, our extension for training with tPIT directly in the waveform domain; and (iii) \textbf{tPIT-latent}, another extension we propose for training Conv-TasNet with tPIT in a learned latent space.

\begin{figure}[t]
    \centering
    \includegraphics[scale=0.395]{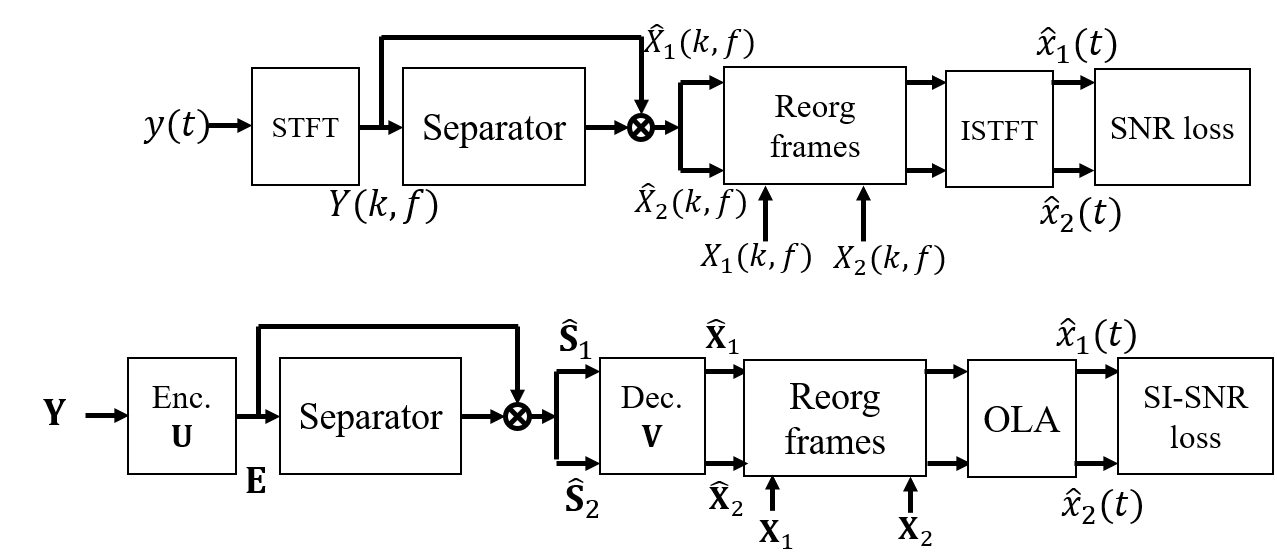}
    \vspace{-6mm}
    \caption{tPIT training for spectrograms (top) and waveforms (bottom).}
    \label{fig:tpit_time}
    \vspace{-4mm}
\end{figure}

\textbf{tPIT-STFT} (Fig. \ref{fig:tpit_time}, top) --- In Deep CASA, the mixture signal $y(t)$ is converted to a complex-valued STFT $Y(k,f)$ where $k$, $f$ denote the discrete frame and frequency indices. A separator, based on Dense-UNet, computes the separated complex STFT $\hat{X}_c(k,f)$ by predicting a multiplicative mask for each speaker $c$. Next, for each individual frame, given the ground truth STFT, the best permutation with the smallest spectral $L_1$ distance is found and used to reorganize frames into speaker-consistent separations.
After the inverse STFT, the signal-to-noise {ratio (SNR) loss is used for training.}

\textbf{tPIT-time} (Fig. \ref{fig:tpit_time}, bottom) --- For adapting tPIT to the waveform domain, we first simply compute the tPIT loss directly in the time domain.
In this paper, we investigate Conv-TasNet~\cite{luo2019conv}, which employs a learned encoder/decoder instead of fixed STFT.
The encoder projects the mixture waveform into a latent space:
\mbox{$\mathbf E = \mathbf U\mathbf Y$}, where \mbox{$\mathbf Y \in \mathbb{R}^{L \times K}$} stores $K$ overlapping frames of length $L$ and stride $S$, \mbox{$\mathbf U \in \mathbb{R}^{N \times L}$} contains $N$ learnable basis filters, and \mbox{$\mathbf E \in \mathbb{R}^{N \times K}$} is the latent space representation of the mixture waveform. The decoder performs the inverse mapping as:
$\mathbf{\hat X}_c = \mathbf{\hat S}^T_c\mathbf V$, where \mbox{$\mathbf V \in \mathbb{R}^{N \times L}$} contains $N$ decoder basis filters, $\mathbf {\hat S}_c \in \mathbb{R}^{N \times K}$ is the $c$th output of the separator, and $\mathbf{\hat X}_c \in \mathbb{R}^{K \times L}$ contains $K$ frames. In this work, we use $L=16$ and $S=8$ samples for our 8 kHz experiments (and $L=32$ and $S=16$ for 16 kHz experiments) and $N=512$.
Consecutive frames in $\mathbf{\hat X}_c$ may not belong to the same speaker due to permutation ambiguity. Accordingly, during training, the best tPIT permutation $\pi^*_k$ for each frame $k$ can be computed independently by the $L_1$ distance in the waveform domain:
\vspace{-2.5mm}
\begin{equation*}
   \pi^*_k=\argmin_{\pi_k\in P}\sum_{c=1}^{C}\left\lvert\mathbf{\hat x}_{c,k}-\mathbf{x}_{\pi_k(c),k}\right\rvert
   \vspace{-0.1mm}
\end{equation*}
where $P$ is the set of all $C!$ permutations, $\mathbf{\hat x}_{c,k}$ is the $k$th frame of the $c$th separator output, and $\mathbf{x}_{\pi_k(c),k}$ denotes the same frame of the ground truth source tied with the $c$th estimate under permutation $\pi_k$. 
After reordering frames according to $\pi^*_k$, overlap-and-add (OLA in Fig. \ref{fig:tpit_time}) is used to reconstruct speaker-consistent predictions. Finally, the scale invariant SNR ({SI-SNR})~\cite{le2019sdr} is used to train the model.

\begin{figure}[t]
    \centering
    \includegraphics[scale=0.4]{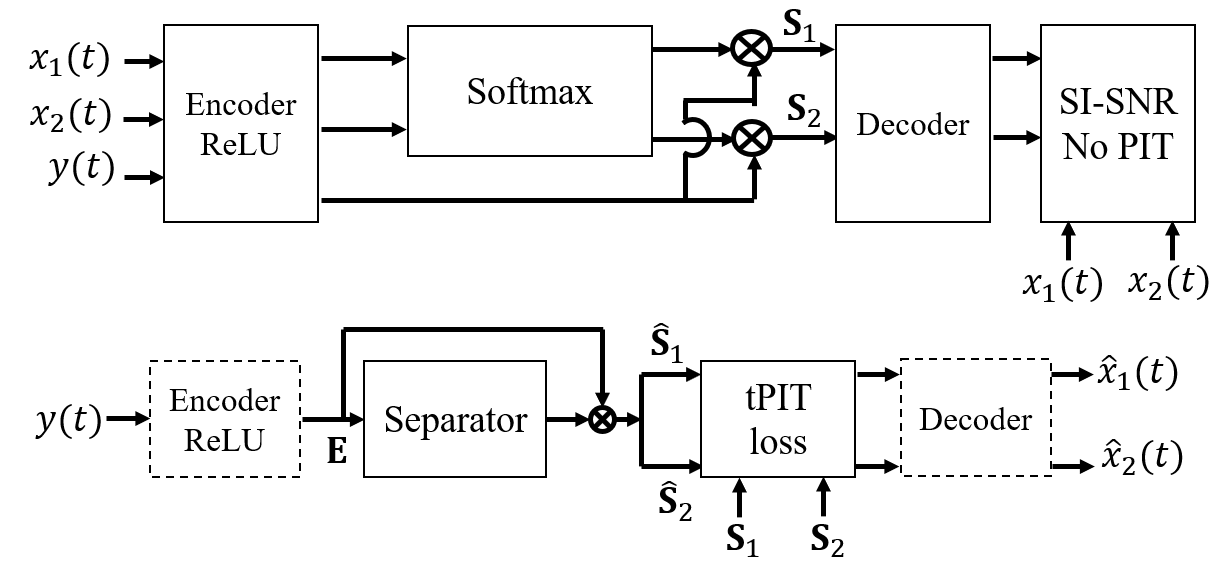}
    \vspace{-4mm}
    \caption{tPIT training in the latent space. First, train the encoder/decoder to generate the optimal latent representation (top). Next, train the separator only with tPIT loss (bottom).}
    \label{fig:tpit_latent}
    \vspace{-4mm}
\end{figure}

\textbf{tPIT-latent} (Fig. \ref{fig:tpit_latent}) --- TasNet-like architectures typically use much shorter frame length and stride than those of spectrogram-based models.
Such short frames may not be adequate to accurately perform tPIT directly in the time domain. Thus, we propose to perform tPIT in a learned latent space. However, due to its latent nature, it's hard to define a pre-existing target for tPIT training. This is a major difference when comparing our work with Deep CASA, which naturally uses STFT as the ground truth.
To overcome this challenge, we train the encoder/decoder and the separator separately. In Fig. \ref{fig:tpit_latent} (top), the encoder converts the mixture signal $y(t)$ and each speech source $x_c(t)$ to a latent space (with non-negative values after the ReLU). The softmax computes the ideal masks from the ground truth signals, which are multiplied with the mixture representation to obtain the ideal latent features $\mathbf{S}_c \in \mathbb{R}^{N \times K}$ of the separated signals, where $N$ and $K$ are the number of basis filters and frames respectively. The decoder performs the inverse transform. The encoder and decoder are learned by minimizing the \text{SI-SNR} loss. Note that, in this step, the ground truth and the separated signals have the same order. Hence, no permutation ambiguity is introduced, which is important for $\mathbf{S}_c$ to become targets when later training the separator.
The idea of training the separator separately from the encoder/decoder was first introduced in~\cite{tzinis2020two}. Our contribution is to use it for setting a visible target for tPIT-latent. In Fig. \ref{fig:tpit_latent} (bottom), the encoder/decoder is frozen (dashed line), and the separator is trained by minimizing the tPIT loss as:
\vspace{-2mm}
\begin{equation*}
   loss_{tPIT}=\frac{1}{KNC}\sum_{k=1}^{K}\min_{\pi_k\in     P}\sum_{c=1}^{C}\left\lvert\mathbf{\hat s}_{c,k}-\mathbf{s}_{\pi_k(c),k}\right\rvert
   \label{eq:tpit_latent}
\vspace{-1mm}
\end{equation*}
\noindent where $\mathbf{\hat s}_{c,k}$ is the $k$th frame (column) of the $c$th predicted source in the latent space $\mathbf{\hat S}_c$, which is compared, by the $L_1$ distance, with the $k$th frame of the pre-trained representation of source $\pi_k(c)$ where $\pi_k$ is a frame level permutation from the set $P$.

\vspace{-2mm}
\subsection{Clustering step}

Since tPIT separates speakers for each frame independently, the permutation frequently changes across frames in the separated signals. The goal of the clustering stage is to predict the permutation for each frame. 
The original clustering step by Deep CASA is based on a pairwise similarity loss in which the cosine distance between every pair of frames, projected in an embedding space, is computed to match either 1 or 0 (permute or not). 
Due to the small frame stride used by Conv-TasNet, this loss can be memory and computationally intensive: $O(K^2)$, where $K$ is the number of frames. We propose a much more efficient solution based on the generalized end-to-end (GE2E) loss~\cite{wan2018generalized}. The latent
features of the mixture $\mathbf E$ and the separated signals $\mathbf {\hat S}_1$ and $\mathbf {\hat S}_2$ in Figs. \ref{fig:tpit_time}, \ref{fig:tpit_latent} are jointly fed through a similarity model, that yields an embedding vector that represents the permutation of each frame. The similarity model is trained such that frames having the same permutation are projected into clusters, enforced by a similarity loss. 
For each frame, we compute its squared Euclidean distance to the mean vectors of all clusters. Then, a permutation softmax (classification) loss is minimized as:
\vspace{-2mm}
$$
   loss_{GE2E}=\sum_{k=1}^K -\log\frac{\exp{-d(\mathbf{h}_{k,p}, \mathbf{m}_p)}}{\sum_{i=1}^{C!}\exp{(-d(\mathbf{h}_{k,p}, \mathbf{m}_i))}}
   \label{eq:ge2e}
   \vspace{-1mm}
$$
\noindent where $\mathbf{h}_{k,p}$ is the embedding of the $k$th frame with a permutation label $p$ (with $p$ obtained from the tPIT stage), $\mathbf{m}_i$ is the mean of the $i$th cluster, and $d(\mathbf{x}, \mathbf{y})=w{\lVert \mathbf{x}-\mathbf{y}\rVert}^2+b$ is the squared Euclidean distance with learnable scale $w>0$, not to change the sign of $d(\cdot)$, and bias $b$.
The GE2E loss was originally developed to train speaker-ID models~\cite{wan2018generalized}, but we adopt it to solve a new problem: to train a permutation similarity model with approximately linear complexity (since $C!\ll K$). Another change is the Euclidean distance, which works better than the cosine distance used for the original GE2E. At test time, K-means (also based on Euclidean distance) is used to cluster the embedding. Finally, the separated latent frames are reordered based on the predicted permutations.

\vspace{-1mm}

\section{\lowercase{u}PIT + PASE for Conv-TasNet}
\vspace{-1mm}

\label{sec:upit+pase}
\begin{figure}[t]
\vspace{-2mm}
    \centering
    \includegraphics[scale=0.4]{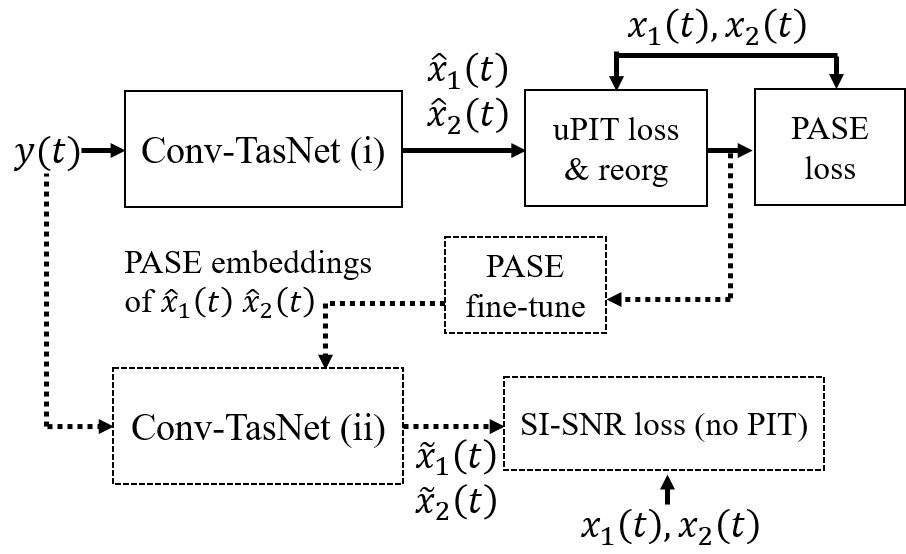}
    \vspace{-4mm}
    \caption{uPIT+PASE (solid) and its cascaded extension (dashed).}
    \label{fig:upit_pase}
    \vspace{-4mm}
\end{figure}
Nachmani et al.~\cite{nachmani2020voice} augmented uPIT by a deep feature loss on speaker embeddings extracted from a pre-trained speaker-ID network and achieved less permutation errors. We also explore this direction, but replacing the speaker-ID deep feature loss with PASE \mbox{(see Fig. \ref{fig:upit_pase})}. The PASE~\cite{pascual2019learning, ravanelli2020multi} encoder was pre-trained in a self-supervised fashion with objectives including speaker-ID, pitch, MFCCs, among others. Hence, PASE embeddings may contain additional relevant information, other than just speaker-ID. Also, the PASE encoder is fully differentiable. Thus, we can construct the uPIT+PASE loss as:
\vspace{-2mm}
$$
loss = uPIT+\sum_{c=1}^{C}{\lVert PASE(\hat{x}_c(t))-PASE(x_{\pi^*_u(c)}(t))\rVert}^2
\label{eq:pase}
\vspace{-2mm}
$$
\noindent where $PASE(\cdot)$ denotes a sequence of PASE embeddings, and  $x_{\pi^*_u(c)}(t)$ is the reference signal tied to the $c$th estimated source under the best uPIT permutation. 
The PASE loss term enforces the frame permutations to align with the best utterance permutation $\pi^*_u$.

Previous works also show that conditioning source separation models with additional features improves performance~\cite{takahashi2020improving, gu2019end, tzinis2020improving}, but whether feature conditioning helps reducing permutation errors has yet to be confirmed. Towards this end, we extend the single-stage uPIT+PASE paradigm to a two-stage cascaded system. The idea is that stage (i) separates the sources, and stage (ii) uses those estimates to generate a conditioning for Conv-TasNet (with PASE features). 
As depicted in Fig. \ref{fig:upit_pase}, where dashed lines denote stage (ii), stage (i) separates the speech sources using a Conv-TasNet trained with uPIT+PASE, and stage (ii) uses PASE (that can now be fine-tuned) to extract features to condition a Conv-TasNet trained from scratch.
Note that the Conv-TasNet of stage (ii) is guided by the separations from the Conv-TasNet in step (i). Given that uPIT already finds the best permutation in stage (i), no PIT is used for training in stage~(ii).


%

 \vspace{-2mm}
 
\section{Experiments}

\vspace{-1mm}

\label{sec:experiments}
We trained various models on the commonly used WSJ0 2-speaker (WSJ0-2mix) database~\cite{merlscript,hershey2016deep}. The training set was created by randomly mixing utterances from 100 speakers at randomly selected SNRs between 0 and 5 dB. 
Previous works found that models trained on WSJ0-2mix might not generalize well to other datasets~\cite{kadiouglu2020empirical,cosentino2020librimix}. To also evaluate model generalization, we tested our models not only on the WSJ0-2mix test set (16 unseen speakers), but also on the recently released Libri-2mix (40 speakers) and VCTK-2mix (108 speakers) test sets~\cite{cosentino2020librimix}. We used the clean version of those datasets, down-sampled at 8 kHz. For our uPIT+PASE experiments, we used the 16 kHz version since PASE was designed to work at this sampling rate. We trained all models with 4 sec speech segments. For the models trained with SI-SNR, we pre-processed the target signals by variance normalization using the standard deviation of the mixture as in~\cite{tzinis2020two}.
As a separator for Conv-TasNet models we used the TCN version by Tzinis et al.~\cite{tzinis2020two}. We used the ADAM optimizer with a learning rate of 1e-3, and divided the learning rate by 2 after 5 consecutive epochs with no reduction in validation loss.

\subsection{Evaluation metrics}

Our metrics measure: the quality of the separation (SI-SNRi), the percentage of frame permutation errors (FER), and the percentage of ``hard" examples (HSR) with an SI-SNRi less than 5 dB. FER and HSR metrics provide insights for studying permutation errors. SI-SNRi: the higher the better. FER and HSR: the lower the better. 

\noindent \hspace{2mm}\textbf{-- SI-SNRi (dB)}: the scale-invariant signal-to-noise ratio improvement measures the degree of separation of different models~\cite{le2019sdr}.

\noindent \hspace{2mm}\textbf{-- FER (\%)}: frame error rate~\cite{liu2019divide}. For each utterance, we count the minimum percentage of inconsistent frame permutations\footnote{The best frame permutation is the one with the smallest spectral $L_1$ distance, to be consistent with Deep CASA. We did not exclude silent frames.} with respect to all possible utterance-level permutations (permute or not).

\noindent \hspace{2mm}\textbf{-- HSR (\%)}: hard-sample rate~\cite{zeghidour2020wavesplit}. HSR measures the percentage of  ``hard" samples (with an SI-SNRi less than 5 dB). Informal listening reveals that these samples contain many speaker swaps. We adjusted the 5dB threshold so that it reflected permutation errors.

\begin{table}[h]
\centering
\vspace{-3mm}
  \caption{SI-SNRi results of the tPIT+clustering algorithms. The ``optimal clusters" use the target signals to reorder frames.}
\vspace{2mm}  
 \begin{tabular}{|l||c|c|c|}
  \hline
 \textbf{} & \textbf{WSJ0} & \textbf{Libri} & \textbf{VCTK}\\
 \hline \hline
 uPIT-waveform & 15.9 & 10.4 & 9.4 \\
 uPIT-STFT & 15.5 & 11.4 & 12.7 \\
 \hline 
 tPIT-STFT + optimal clusters & 18.5 & 16.0 & 15.5\\
 tPIT-STFT + clustering & 17.5 & 13.9 & 13.6\\
 \hline
 tPIT-time + optimal clusters & 16.7 & 12.1 & 13.0\\
 tPIT-time + clustering & 15.5 & 9.8 & 9.9\\
 \hline
 tPIT-latent: enc/dec (Fig. 2, top) & 55.5 & 54.9 & 53.9\\
 tPIT-latent + optimal clusters & 17.6 & 12.9 & 13.7\\
 tPIT-latent + clustering & 16.5 & 11.0 & 11.0\\
 \hline  \hline
 \multicolumn{3}{|l}{\textbf{tPIT-latent + clustering: clustering loss variants}} & \multicolumn{1}{c|}{}\\
 \hline  \hline
 pairwise similarity loss  & 16.2 & 10.7 & 10.8 \\
 GE2E loss  & 16.5 & 11.0 & 11.0 \\
 \hline

 \end{tabular}

 \vspace{-4mm}
 \label{tpit_tasnet}
\end{table}

\begin{table*}[t]
\centering
\vspace{-10mm}
 \caption{SI-SNRi and FER of uPIT+PASE and its cascaded version for Conv-TasNet. Results with speech at 16kHz.}
 \vspace{2mm}
 \begin{tabular}{|c||c|c||c|c||c|c||c|c|}
  \hline
 {} & \multicolumn{2}{c||}{\textbf{uPIT-waveform}} & \multicolumn{2}{c||}{\textbf{uPIT+PASE}} & \multicolumn{2}{c||}{\textbf{uPIT+PASE cascaded}} &
 \multicolumn{2}{c|}{\textbf{tPIT-latent+clustering}}\\
 \cline{2-9}
 {} & SI-SNRi & FER & SI-SNRi & FER & SI-SNRi & FER & SI-SNRi & FER\\
 \hline  \hline
 WSJ0 & 15.5 & 5.2 & 15.9 & 4.5 & 17.5 & 4.6 & 16.0 & 4.3\\ 
 \hline
 Libri & 10.7 & 9.0 & 10.8 & 8.0 & 11.9 & 7.6 & 11.1 & 7.8\\
 \hline
 VTCK & 9.5 & 12.4 & 9.9 & 11.2 & 10.9 & 11.3 & 10.9 & 9.5 \\
  \hline

 \end{tabular}
\vspace{-4mm}
 \label{pase_results}
\end{table*}

\subsection{\lowercase{t}PIT+clustering discussion}

\begin{table}[h]
\centering
\vspace{-3mm}
 \caption{Error analysis results: tPIT+clustering.}
\vspace{2mm}
 \begin{tabular}{|c||c|c|c|c|c|c|}
 \hline

 {} & \multicolumn{2}{c|}{\textbf{uPIT}} & \multicolumn{2}{c|}{\textbf{tPIT-latent }} &
 \multicolumn{2}{c|}{\textbf{tPIT-STFT}}\\
  {} & \multicolumn{2}{c|}{\textbf{(waveform)}} & \multicolumn{2}{c|}{\textbf{+ clustering}} &
 \multicolumn{2}{c|}{\textbf{+ clustering}}\\
 \cline{2-7}
 {} & FER & HSR & FER & HSR & FER & HSR \\
 \hline \hline
 WSJ0 & 6.1 & 6.0 & 5.4 & 1.8 & 4.9 & 2.2 \\
 \hline \
 Libri & 9.4 & 14.8 & 8.5 & 9.1 & 6.6 & 7.4 \\
 \hline 
 VTCK & 12.3 & 22.8 & 9.4 & 10.7 & 7.8 & 7.2\\
 \hline

 \end{tabular}
\vspace{-4mm}
 \label{error_rate}
\end{table}

In our clustering model, the latent features $\mathbf E$ and $\mathbf {\hat S}_1$, $\mathbf {\hat S}_2$ first go through a 1-D batch normalization, and are concatenated and linearly projected to 512 channels, from which two 1-D convolutional layers (kernel size 3 with PReLU) further process those. The rest of the model follows the TCN architecture of the clustering model in Deep CASA~\cite{liu2019divide}. Finally, a linear layer generates a 40-dim embedding (whose $L_2$ norm is normalized to 1).

\begin{figure}[h]
    \centering
    \vspace{-4mm}
    \includegraphics[scale=0.48]{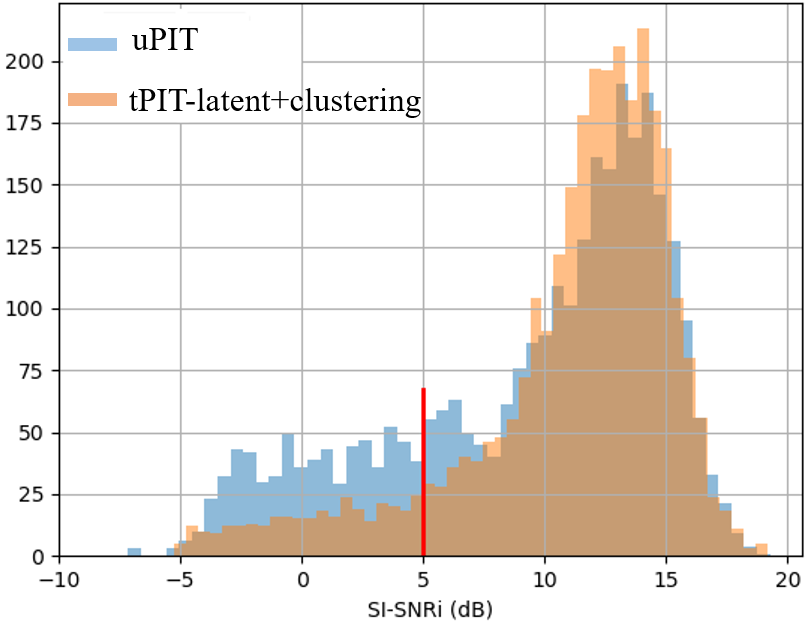}
    \vspace{-4mm}
    \caption{Histogram: SI-SNRi (dB) results of tPIT-latent+clustering on VCTK-2mix. Red line: 5dB threshold defining ``hard" samples.}
    \label{fig:histogram}
    \vspace{-3mm}
\end{figure}

\begin{itemize}

\item \textbf{Spectrogram vs. waveform-based models for tPIT+clustering}. In Table 1, we compare Conv-TasNet models with two STFT-based models: uPIT-STFT (the uPIT-trained variant of Deep CASA), and tPIT-STFT+clustering (Deep CASA). We trained Deep CASA using the official code~\cite{deepcasa_code}. 
Note that spectrogram-based models generalize better than waveform ones. 
It's known that models using signal processing transforms are less prone to overfitting than those using learnable encoders/decoders~\cite{ditter2020multi}, but the optimal encoder/decoder results for tPIT-latent (obtained following the training procedure depicted in Fig.2, top) suggest that the learned encoder/decoder does not overfit. Instead, it reveals that the separator can be a key factor for model generalization.

\hspace{4mm} Further, when looking at the error analysis in Table 3, note that tPIT-STFT+clustering (Deep CASA) FER scores are the lowest.
Deep CASA uses much longer frames (32 ms) than that of Conv-TasNet (2 ms), which might enable more accurate frame-based speaker separation (indicated by the tPIT-STFT+optimal clusters result in Table 1). 
Although short frames seem to be critical for achieving state-of-the-art results with waveform models~\cite{luo2020dual}, they also pose challenges related to permutation errors. These results denote that Deep CASA (a STFT-based model) is effective at reducing permutation ambiguity and improving generalization.

\item \textbf{tPIT+clustering for Conv-TasNet}. In Table 1, we study the tPIT+clustering extensions we propose for Conv-TasNet: tPIT-time and tPIT-latent.
The uPIT-waveform baseline does not generalize, note the large performance drop on the other two datasets. Further, tPIT-latent+clustering achieves improvements that generalize.
However, tPIT-time does not yield as good performance as tPIT-latent, showing the advantage of tPIT in the latent space.

\hspace{4mm} Table \ref{error_rate} depicts our error analysis, where tPIT-latent+clustering outperforms uPIT-waveform. Fig. \ref{fig:histogram} shows the histogram of the VCTK-2mix SI-SNRi results. The tPIT-latent+clustering method pushes the distribution towards the right, significantly improving on ``hard" examples (with SI-SNRi $<$ 5 dB).

\item \textbf{GE2E loss}. To study the GE2E loss we propose, we trained the same tPIT-latent+clustering model with the pairwise similarity loss in Deep CASA. Since their loss is memory intensive, to fit into GPU memory, for each forward pass, we compute and back propagate the loss 4 times (each over a 1 sec segment). Besides being memory and computationally efficient, the GE2E loss also provides better results (Table 1).

\end{itemize}

\vspace{-3mm}
\subsection{\lowercase{u}PIT+PASE discussion}
 We used the PASE+ model~\cite{pase_code} pre-trained with 50 hours of LibriSpeech (2338 speakers). It generates a 256-dim embedding every 10 ms. In the second stage of our cascaded system, each depthwise convolution in Conv-TasNet's separator is conditioned with the PASE features of the separated speakers through a FiLM~\cite{perez2018film} layer. The weights in FiLM are shared across all layers, since this yielded better results than using dedicated FiLM weights per layer.

The results of our experiments are listed in Table \ref{pase_results}. Adding the PASE term to uPIT (single stage) improves SI-SNRi and reduces permutation error over all test sets. However, its improvements are smaller than those obtained by the tPIT-latent+clustering approach, implying the advantage of explicitly learning to permute with the clustering step. 
Also, note that the cascaded system we propose improves SI-SNRi across all test sets, but this is not the case for FER. This result suggests that the cascaded approach does not help amending the permutation errors introduced by uPIT in the first-stage. 

\vspace{-2mm}
\subsection{Additional discussion: previous results on WSJ0-2mix}
\vspace{-1mm}

Wavesplit~\cite{zeghidour2020wavesplit}, a recent WaveNet-like model, improved SI-SNRi to 21 dB by combining tPIT+clustering and speaker conditioning, denoting the potential of these two directions that we also find promising. It also reduced HSR (based on a 10 dB threshold) to 5.6\%. We also measured the ``10 dB HSR" for our tPIT-latent+clustering Conv-TasNet and obtained 5.0\%---the 5 dB threshold was used in previous results to better reflect permutation errors. 

Prob-PIT~\cite{yousefi2019probabilistic} considers the probabilities of all utterance level permutations, rather than just the best one, improving the initial training stage when wrong alignments are likely to happen.
A similar idea is employed by Yang et al.~\cite{yang2020interrupted}, who trained a Conv-TasNet with uPIT and fixed alignments in turns, reporting 17.5 dB SI-SNRi. They also implemented Prob-PIT for Conv-TasNet and obtained \mbox{15.9 dB}.
Our SI-SNRi results (16.5 and \mbox{17.5 dB}) are promising. Further, we also confirmed the reduction of permutation errors and generalization of improvements to other test sets, which was not tested in~\cite{yousefi2019probabilistic, yang2020interrupted}.

\vspace{-2mm}

\section{Conclusion}
\vspace{-1mm}
\label{sec:conclusion}

We explored two PIT directions to tackle the permutation ambiguity problem: tPIT+clustering and uPIT+PASE.
tPIT+clustering was originally proposed in the STFT domain by Deep CASA. We adapted it to work with Conv-TasNet, a waveform-based model. Our extensions include tPIT-time, optimizing tPIT on the waveform domain, and tPIT-latent, optimizing tPIT over a learned latent space.
{Further}, the original clustering model by Deep CASA does not scale for waveforms, and we propose using the GE2E loss that is efficient and obtains better results.
Our results on training Conv-TasNet with tPIT+clustering show that tPIT-latent outperforms tPIT-time and uPIT: it generalizes to other test sets and obtains less permutation errors.
We also found that, although uPIT+PASE reduced permutation errors, it was not as effective as tPIT+clustering. 
Also, tPIT-STFT+clustering (Deep CASA) is more effective at improving generalization and reducing permutation ambiguity than the studied waveform-based models. 
This result provides a new perspective when comparing spectrogram- and waveform-based models: spectrogram-based models can help reducing permutation errors.


\bibliographystyle{IEEEbib}
\bibliography{z_bibliography}
\end{document}